# 60 GHz High Data Rate Wireless Communication System

Lahatra Rakotondrainibe, Yvan Kokar, Gheorghe Zaharia, and Ghaïs El Zein
IETR-INSA, UMR CNRS 6164, 20, Av. des Buttes de Coësmes, CS 14315, 35043 Rennes Cedex, France
lrakoton@insa-rennes.fr; yvan.kokar2@insa-rennes.fr; gheorghe.zaharia@insa-rennes.fr; ghais.el-zein@insa-rennes.

*Abstract* – **This paper presents the design and the realization of a 60 GHz wireless Gigabit Ethernet communication system. A differential encoded binary phase shift keying modulation (DBPSK) and differential demodulation schemes are adopted for the IF blocks. The Gigabit Ethernet interface allows a high speed transfer of multimedia files via a 60 GHz wireless link. First measurement results are shown for 875 Mbps data rate.**

## I. INTRODUCTION

During the last decade, substantial knowledge about the 60 GHz millimeter-wave (MMW) channel has been accumulated and different architectures have been analyzed to develop new MMW communication systems for commercial applications [1-2]. Due to the large propagation and penetration losses, 60 GHz WLANs (Wireless Local Area Networks) are primarily intended for use in short-range and single room environments. Moreover, demands for high-speed multimedia data communications, such as a huge data file transmission and real-time video streaming, are markedly increasing. Hence, one of the most promising solutions to achieve a gigabit class wireless link is to use millimeter-waves for the carrier frequency. Many system proposals under the IEEE 802.15.3c Task Group for wireless local personal network (WPAN) have been studied [3]. The assignment of a large unlicensed bandwidth (5 to 7 GHz) around 60 GHz created new opportunities for 60 GHz front-end technology. High frequency and even MMW analog communication circuits, which were traditionally built on more expensive technologies such as bipolar or Gallium Arsenide (GaAs), are gradually implemented on CMOS [4]. Also, recent progresses in electronic devices have opened the way to operate at data rate up to several Gbps.

This paper presents the design and implementation of a high data rate wireless communication system operating at 60 GHz developed at IETR. The rest of this paper is organized as follows. Section II describes the wireless communication system used for Gigabit Ethernet applications. The transceiver architecture is described in section III. Section IV presents the baseband modules architecture and data format. Results of our experiments are shown in section V. Finally, conclusions and future work are drawn in Section VI.

This work is part of the research project Techim@ges supported by the French "Media & Network Cluster" and the COMIDOM project supported by the "Région Bretagne".

## II. WIRELESS COMMUNICATION SYSTEM

The wireless communication system is composed of radio frequency (RF) blocks, intermediate frequency (IF) blocks and baseband (BB) blocks. The Gigabit Ethernet interface is used to connect a home server to a wireless link with around 800 Mbps bit rate. Fig. 1 and Fig. 2 respectively show the wireless Gigabit Ethernet transmitter and receiver.

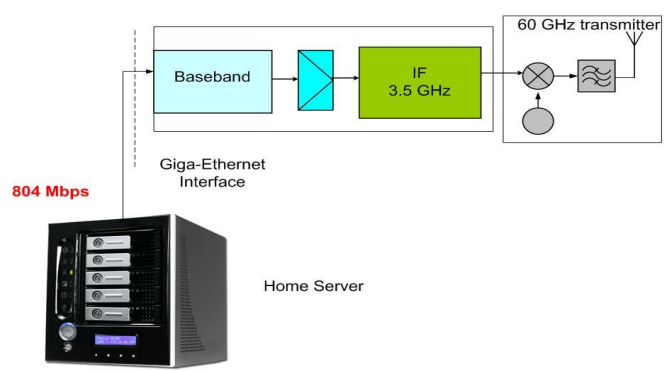

Figure 1. Wireless Gigabit Ethernet transmitter

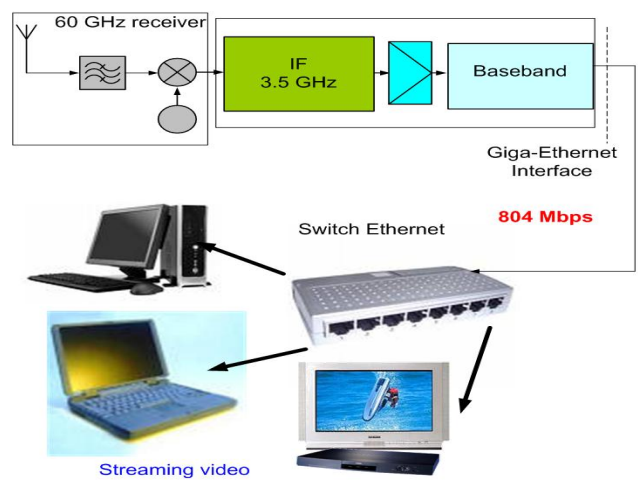

Figure 2. Wireless Gigabit receiver

The first application of the system in a point-to-point link configuration is the streaming video. The system must operate in indoor, line-of-sight (LOS) domestic environments. Due to this application, the wireless communication link should be considered in a single room. Therefore, the Tx-Rx distance does not exceed 10 meters.

### III. TRANSCEIVER BLOCKS

The system has a single carrier transmission scheme based on DBPSK modulation and non-coherent demodulation. This system, compared to higher order constellations or OFDM systems, is more resistant to phase noise and power amplifiers (PAs) non-linearities. OFDM requires large back-off for PAs, high stability and low phase noise for local oscillators. Furthermore, the implementation of such a system is simple. The design and the realization of the RF blocks were described in [4]. These blocks are indicated in gray color in Fig. 3 and Fig. 4. The block diagram of the transmitter is shown in Fig.3.

An automatic gain control (AGC) has 2.5 dB Noise Figure with 20 dB dynamic range. The AGC circuit monitors the output signal and the power detector which controls the gain of the AGC amplifier. When the Tx-Rx distance increases from 1 m to 10 m, the power of the received IF signal decreases from -34 dBm to -54 dBm and the AGC gain varies from 8 dB to 28 dB. A Low Noise Amplification stage (LNA) with a gain of 40 dB is used to achieve sufficient gain. Hence, DBPSK modulation at the transmitter and non-coherent detection at the receiver represent the classical solution for avoiding the problem of phase ambiguity. The differential demodulation method reduces the effect of signal phase variation in the channel if the phase of the carrier frequency does not vary appreciably during two successive symbols.

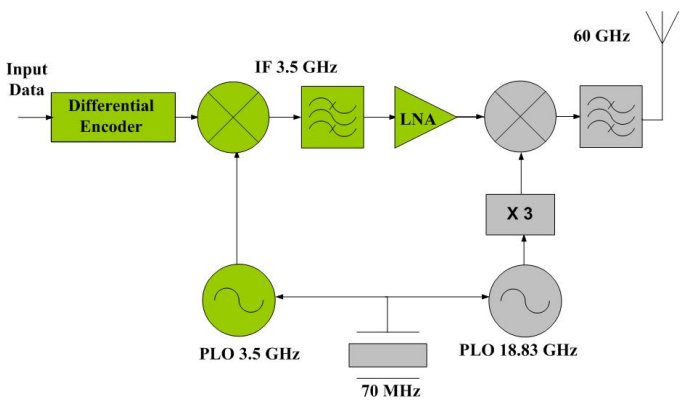

Figure 3. 60 GHz single carrier transmitter

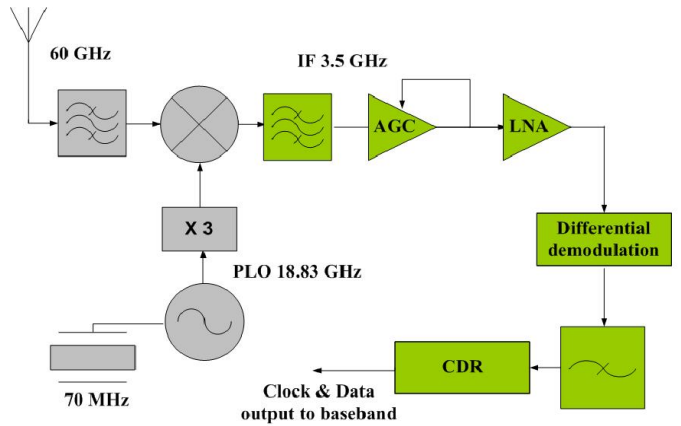

Figure 4. 60 GHz single carrier receiver

The input data are differentially encoded using an exclusive OR (XOR) and a D flip-flop. The differential encoder performs the modulo two addition of the input data bit with the previous output bit. This encoder is implemented using PECL circuits. DBPSK modulation is used to eliminate the need of synchronous recovery of the carrier frequency at the receiver and to simplify the receiver design. The obtained data are used to modulate an IF carrier generated by a 3.5 GHz phase locked oscillator (PLO) with a 70 MHz external reference. The IF modulated signal is then fed to a band-pass filter with a bandwidth of 2 GHz. The IF amplifier has a gain ($S_{21}$) of 16 dB within 2-6 GHz band.

The RF transmitter consists of a mixer ($P_{1dBm}$ = 10 dBm) which up-converts the IF signal to 60 GHz using a 56.5 GHz Local Oscillator (LO) signal. The LO frequency is obtained with an 18.83 GHz PLO with the same 70 MHz reference and a frequency tripler. The phase noise of the 18.83 GHz PLO signal is about – 110 dBc/Hz at 10 kHz off-carrier. The upper sideband is selected using a band-pass filter. The 0 dBm obtained signal is fed to the horn antenna with a gain of 22.4 dBi.

The block diagram of the receiver is shown in Fig. 4. The receive antenna, identical to the transmit antenna, is connected to a band-pass filter (59-61 GHz). The obtained signal is down-converted to the IF signal which is fed to a band-pass receive filter with a bandwidth of 2 GHz.

This simple differential demodulation enables the coded signal to be demodulated and decoded. The mixer conversion loss is around 7 dB and the output power is around 0 dBm. Basically, the demodulator design is based on a mixer and a delay line. In the DBPSK demodulator, a band-pass delay line provides the one bit delay ($\tau$ = 1.14 ns) required for the demodulation. Following the loop, a low-pass filter (LPF) removes the high-frequency signal. The cut-off frequency of the LPF is 1.8 GHz. From the waveform reshaping after the clock and data recovery circuit (CDR), the bit stream is transferred into baseband modules. The CDR automatically locks to the data rate without the need for an external reference clock or programming.

### IV. BASEBAND BLOCKS

The baseband blocks perform the channel coding and decoding in the FPGA Xilinx Virtex 4. Fig. 5 depicts the frame structure. The channel coder is used to add redundancy to the binary information. To achieve frame synchronization in packet-based communication systems, a known sequence (preamble) is sent at the beginning of each packet. The preamble is a pseudo-noise (PN) sequence of 31 bits with one additional bit to provide 4 bytes. One more byte "frame header" should be added besides 4 preamble bytes to assure the scrambling operation, as it will be explained later.

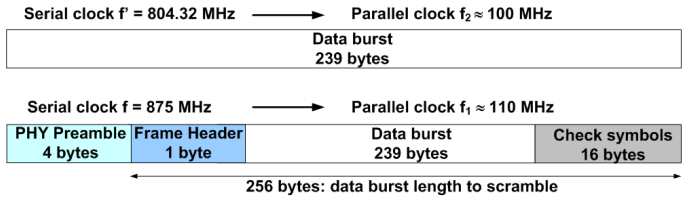

Figure 5. Frame structure

The channel encoder and decoder realized within FPGAs require two different clock frequencies. In our design, these two clocks are obtained by dividing by 8 the frequencies 875 MHz and 804.32 MHz. The 875 MHz frequency is obtained from a 3.5 GHz PLO by using a frequency divider by 4. The 804.32 MHz clock is obtained from the 875 MHz clock through a clock manager circuit programmed by the FPGA. Fig. 6 describes the transmitter baseband block. It consists of a serial to parallel (S/P) conversion, encoding control, Reed-Solomon (RS) encoding, scrambling and parallel to serial (P/S) conversion.

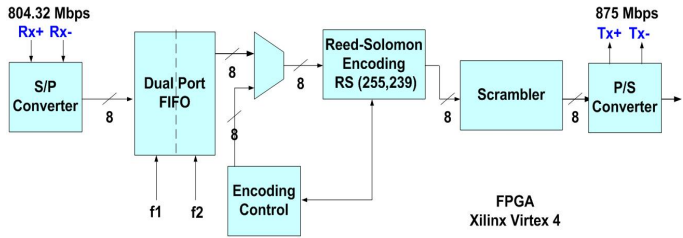

Figure 6. Transmitter baseband block

Due to the maximum frequency supported by the FPGA (around 200 MHz), a parallel encoding must be applied to support 804.32 Mbps data throughput. Hence, the S/P converter parallelizes the input data (804.32 Mbps) in eight bits for the received code word at 100 MHz. The clocks 100 MHz and 110 MHz are respectively obtained from the S/P and P/S converters within the FPGA. The data bytes from the S/P are written into the FIFO dual port at 100 MHz. The FIFO dual port has been set up to use two different frequencies for writing in (f1) and reading out (f2). Therefore, the reading can be started when the FIFO is half full. The encoding control is a module that monitors the Reed-Solomon (RS) encoder and the scrambler. The RS encoder reads one byte every clock period. However, the encoding control indicates that neither coding nor scrambling is applied to 4 preamble bytes. At the next clock edge, on the read port of the FIFO, the byte is read at a higher rate of 110 MHz. After the 239 clocks periods, the encoding control interrupts the bytes transfer during 16 triggered clock periods before retransmitting the preamble sequence. The encoder takes 239 data bytes and appends 16 control bytes to make a code word of 255 bytes. Basically, the scrambling facilitates the work of a clock recovery circuit of the receiver (eliminating long sequences of '0' or '1'). A different PN sequence of 31 bits is convenient completed with one bit in order to obtain a 4 bytes scrambling sequence. In fact, the PN preamble sequence and the PN scrambling sequence form Gold's pair in order to provide a low cross-correlation. This method prevents false detections of the preamble within the scrambled data. The encoding control adds 1 "frame header" byte for the balanced scrambling operation so that the number of 256 data coded bytes (except 4 preamble bytes) is a multiple of 4. The final step consists of an S/P conversion of data bytes. The serial data coded at 875 Mbps are finally transmitted to the differential encoder.

Fig. 7 depicts the diagram of the receiver baseband block. Each byte of the serial input data (875 Mbps) send by the CDR is converted into 8 parallel bits by an S/P converter.

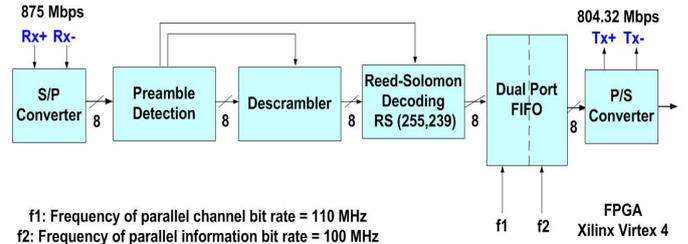

Figure 7. Receiver baseband block

The preamble detection consists of identifying the 4 bytes PHY preamble and the realization of the bytes synchronization as shown in Fig. 8. Due to the periodical repetition of the preamble, in order to obtain a very low false detection probability, two successive preambles must be detected. Hence, two banks of 8 correlators of 32 bits are used. In order to realize the byte synchronization, each correlator must analyze a 1-bit shifted sequence. Therefore, each bank of eight correlators analyzes $32 + 7 = 39$ bits. If the 32 bits preamble is present within 39 bits, only one of the 8 correlators will indicate the convenient byte alignment (synchronization).

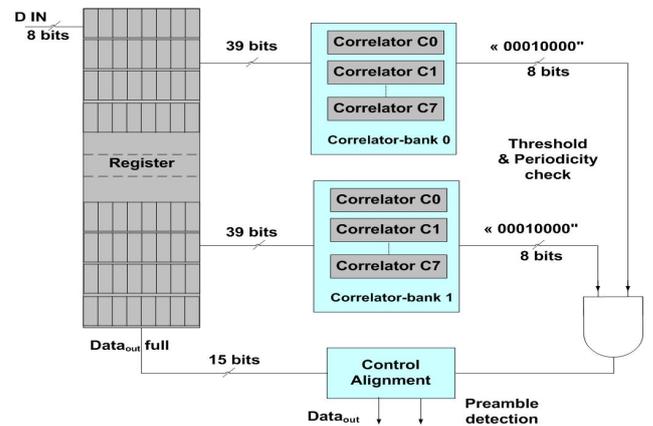

Figure 8. Correlator architecture and control alignment

If, in the both correlator-banks, the detection of PHY preamble is indicated by the same correlator $C_k$, this operation is validated and within 15 output successive bits, 1 byte shifted by k bits is used as a "good" output byte. Therefore, the PHY preamble detection also performs the byte synchronization. Moreover, it controls the descrambler which converts the scrambled data into their original values.

The threshold used by the correlators is chosen in order to obtain the best trade-off between a high value of the detection probability and a very small value of the false detection probability. Furthermore, the preamble detection signals the RS decoder at the beginning of incoming data to be decoded. The RS decoder must operate at 110 MHz. The RS decoder processes each block and attempts to correct errors and recover the original data. The decoder can correct up to 8 bytes that contain errors. The output decoder data are written discontinuously in the FIFO dual port at 110 MHz. Therefore, another frequency around 100 MHz read out continuously the data stored in the register. The serial data decoded at 804.32 Mbps are transmitted into Gigabit Ethernet Interface or to a Bit Error Analyzer after the P/S conversion.

## V. EXPERIMENTAL RESULTS

The measurement of the frequency and impulse responses of RF blocks (Tx/Rx), including LOS channel, is shown in Fig. 9. The main objectives are to determine the available bandwidth for the whole RF system and to estimate the effects of the multipath channel.

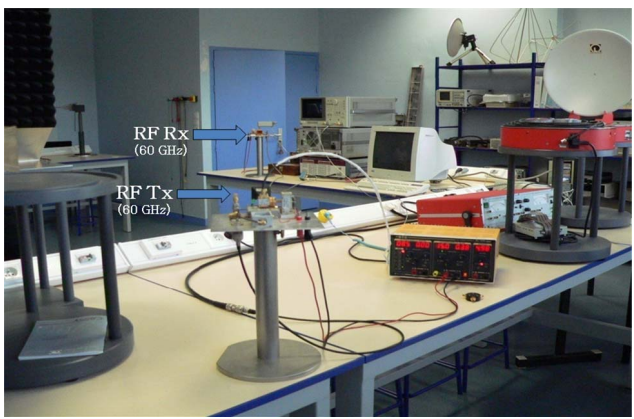

Figure 9. Measurement of RF blocks (Tx & Rx)

Measurements were performed using an HP 8753 D network analyzer. After the calibration of the measurement set-up, a set of channel characteristics can be extracted: frequency and impulse responses, power attenuation and delay spread. As shown in Fig. 10, 2 GHz bandwidth is available. The side lobes presented in Fig. 11 are mainly due to RF components imperfections. Back-to-back tests have been realized and similar results for the frequency and impulse responses were obtained. Directional antennas are essential in 60 GHz band to mitigate the multi-path fading effects. This is acceptable for point-to-point communications links, with minimal multi-path interference, but MMW-WLAN will require equalization techniques or the use of OFDM to overcome multi-path interference while maintaining a high data rate. The remaining significant impact on the system caused by the radio channel is the frequency selectivity which induces intersymbol interference (ISI). However, the use of differential demodulation allows the reduction of ISI effects.

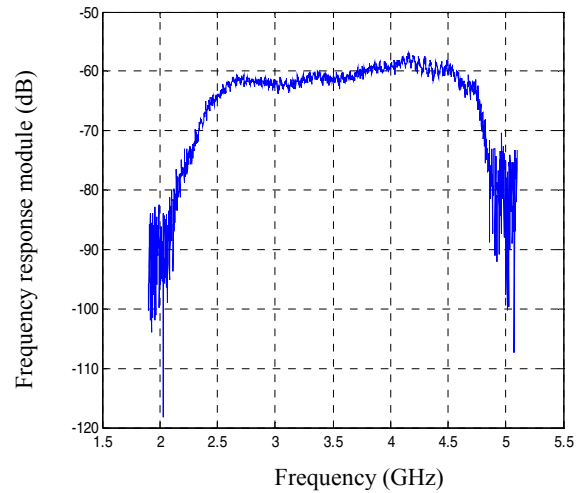

Figure 10. Frequency response of RF blocks, distance 10 m

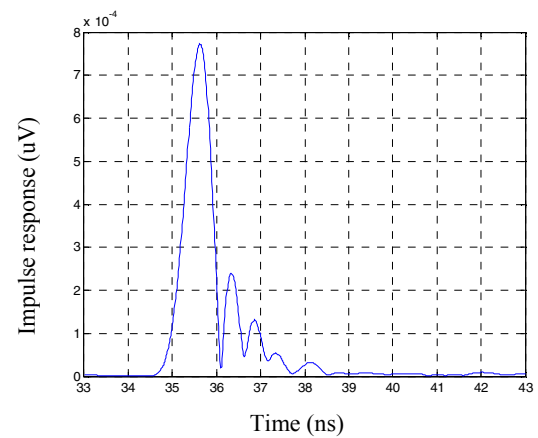

Figure 11. Impulse response of RF blocks, distance 10 m

Table 1 gives the main physical parameters of this 60 GHz wireless communication system.

TABLE I. PHYSICAL PARAMETERS

| Carrier frequency | 60 GHz |
|---|---|
| Modulation scheme | DBPSK, differential demod. |
| Bandwidth | 2 GHz |
| Output Power (EIRP) | 22.4 dBm |
| FEC scheme | RS (255,239) |
| Channel bit rate | 875 Mbps |
| Information bit rate | 804.32 Mbps |
| PHY Preamble | Gold sequence 32 bits |
| Application | File transfer, video streaming |
| Operation range | 10 m in LOS |

The test of IF and RF blocks was realized by using a PN sequence with a length of $2^7-1$ bits generated by HP70841B Pattern Generator. Fig. 12 shows the test of the differential encoder at 875 Mbps. The power spectrum of the IF signal is given in Fig. 13.

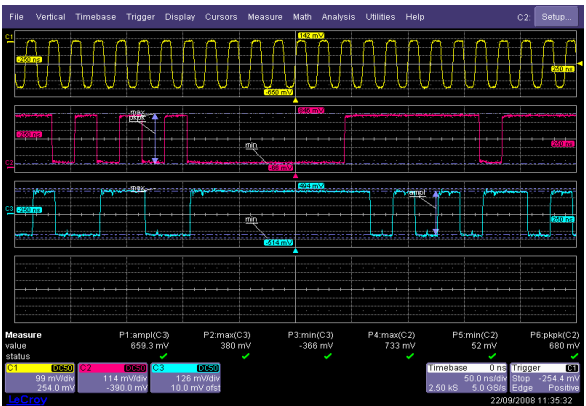
— Input clock   --- Input data   --- Output data
Figure 12. Test of differential encoder at 875 Mbps

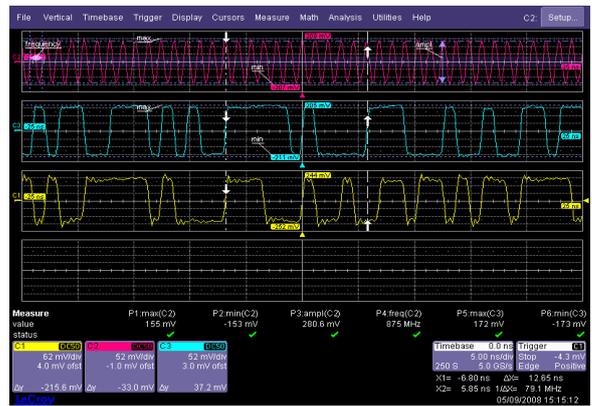
— Input Data   --- Output data   --- Output clock
Figure 15. Measurement on clock and data recovery

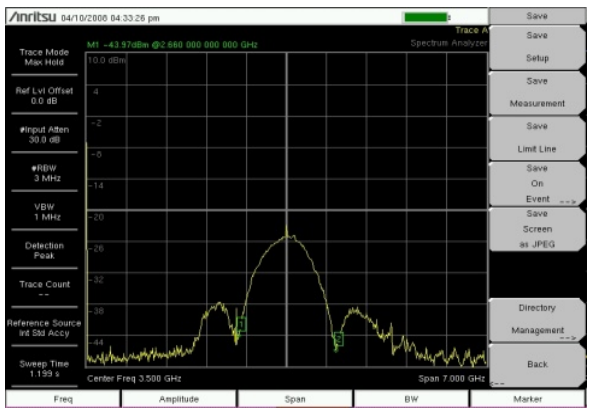
Figure 13. Power spectrum of IF the filtered signal

The eye diagram shown in Fig. 14 (output of the LPF of the differential demodulator) was obtained for 10 m Tx-Rx separation without error correcting coding. This open eye diagram is a suitable indication of the good quality of the data transmission. From this eye pattern, the recovered data can be reached at half period of time and has to be noise resistant.

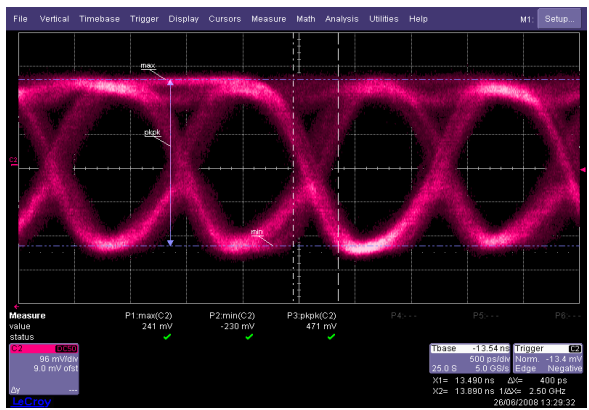
Figure 14. Output eye diagram at 875 Mbps, d(Tx-Rx) = 10 m

Fig. 15 depicts the test of the data and clock recovered after the demodulation. The obtained results show a perfect agreement with the input data.

## VI. CONCLUSION

This paper has described the design and implementation of a high data rate 60 GHz wireless communication system for short range applications. The proposed system provides a good trade-off between performance and complexity. An original method used for the byte synchronization is also described in this paper. This method allows a high detection probability and a very small probability of the false detection.

First measurement results validate the architecture of the whole system. Bit error rate measurements will be soon carried on. An increase of the data rate exceeding 1 Gbps using other modulations (QPSK, OFDM) is planned. Equalization methods are still under study. The demonstrator will be further enhanced to prove the feasibility of Gigabit wireless communications in different configurations.

## ACKNOWLEDGMENTS

The authors specially thank Guy Grunfelder (CNRS engineer) for his technical contributions during the system realization.

## REFERENCES

[1] P. Smulders "Exploiting the 60 GHz Band for Local Wireless Multimedia Access: Prospects and Future Directions", Eindhoven University of Technology, IEEE Communications Magazine, January 2002.

[2] N. Guo, R. C. Qiu, S. S. Mo, and K. Takahashi, "60-GHz Millimeter-Wave Radio: Principle, Technology, and New Results", EURASIP Journal on Wireless Communications and Networking, ID 68253, Sept. 2006.

[3] IEEE 802.15. WPAN Millimeter Wave Alternative PHY Task Group 3c (TG3c), www.ieee802.org/15/pub/TG3c.html

[4] S. Guillouard, G. El Zein, J. Citerne, "Wideband Propagation Measurements and Doppler Analysis for the 60 GHz Indoor Channel", Proceedings of The IEEE MTT-S International Microwave Symposium, Anaheim - CA, USA, June 13-19, 1999, pp. 1751-1754.

[5] L. Rakotondrainibe, G. Zaharia, G. El Zein, Y.Lostanlen. "Indoor channel measurements and communication system design at 60 GHz", XXIX URSRI General Assembly, Chicago, 7-16 August 2008